\documentclass{PoS}

\title{Intrinsic five-quark Fock states of baryons and meson-nucleon $\sigma$-terms}

\ShortTitle{meson-nucleon $\sigma$-terms}

\author{\speaker{C. S. An}\thanks{It is a pleasure to thank the Organizers for partial financial support
to attend the Conference.}\\
        Institute of High Energy Physics, CAS, P. O. Box 918-4, Beijing 100049, China\\
        Theoretical Physics Center for Science Facilities, CAS, P. O. Box 918-4, Beijing 100049, China\\
        E-mail: \email{ancs@ihep.ac.cn}}

\author{B. Saghai\\
        Institut de Recherche sur les
lois Fondamentales de l'Univers, DSM/Irfu, CEA/Saclay, F-91191 Gif-sur-Yvette, France\\
        E-mail: \email{bijan.saghai@cea.fr}}

\abstract{
An extended chiral constituent quark approach, which embodies higher Fock five-quark components
in the baryons wave-functions, is employed to study the sea quark content of the octet baryons. 
We report on the probabilities of $\bar{u}$, $\bar{d}$ and 
$\bar{s}$ in the nucleon, $\Lambda$, $\Sigma$ and $\Xi$ baryons, 
arising from the intrinsic five-quark states in the baryons wave functions.
Based on those probabilities, results for meson-nucleon $\sigma$-terms; 
namely $\sigma_{\pi N}, \sigma_{K N}$ and $\sigma_{\eta N}$, are presented.}

\FullConference{Sixth International Conference on Quarks and Nuclear Physics,\\
		April 16-20, 2012\\
		Ecole Polytechnique, Palaiseau, Paris}

\begin{document}

\section{Introduction}
\label{intro}
%
%
The pion-nucleon nucleon sigma term, $\sigma _{\pi N}$, can be considered as a test of QCD,
where the chiral-symmetry breaking is due to representation of $SU(3)\times SU(3)$; 
see e.g. Refs.~\cite{Huang:2007dy,Durr:2011mp,Bali:2011ks,Shanahan:2012wh} and references 
therein.

A thorough investigation of baryon's structure involves nonpertubative QCD.
Moreover, the strange quark component, a purely vacuum polarization effects, 
plays a crucial role in that realm. 
In a series of recent papers~\cite{An:2010wb,An:2011sb,An:2012kj} we have studied contributions
from the genuine five-quark components in baryons to various properties of 
baryons~\cite{An:2010wb,An:2011sb} and have determined~\cite{An:2012kj} the probabilities of five-quark 
components ($u \bar u$, $d \bar d$ and $s \bar s$) in ground state octet baryons.
In this paper we extend our studies to the meson-nucleon $\sigma$-terms.

Our formalism is based on an extended chiral constituent quark model ($E \chi CQM$) approach and embodies 
all possible five-quark mixtures in the octet baryons wave-functions.
Using our approach, we put forward predictions on the probabilities of $u \bar u$, $d \bar d$ and $s \bar s$
in ground state baryons.

The present manuscript is organized in the following way: in section~\ref{Theo}, we briefly present  
the theoretical frame.
Numerical results are given in section~\ref{Res}, putting forward predictions for the probabilities of 
different five-quark configurations in the nucleon, $\Lambda$, $\Sigma$ and $\Xi$ baryons and extract the 
meson-nucleon $\sigma$-terms.
Finally, section~\ref{Sum} contains summary and conclusions.

%
%
\section{Theoretical frame}
\label{Theo}
In the extended constituent quark model~\cite{An:2012kj}, wave function for a baryon is expressed as
\begin{equation}
 |\psi\rangle_{B}=\frac{1}{\mathcal{\sqrt{N}}}\left[|QQQ\rangle+
 \sum_{i,n_{r},l}C_{in_{r}l}|QQQ(Q\bar{Q}),i,n_{r},l\rangle \right]\,,
\label{wfn}
\end{equation}
where the first term is the conventional wave function for the baryon with three
constituent quarks, and the second term is a sum over all possible higher Fock
components with a $Q\bar{Q}$ pair.
Here we denote light quark-antiquark pair as $Q\bar{Q}\equiv q\bar{q}$ (with $q \equiv u,~d,$)
and strange quark-antiquark pairs as $Q\bar{Q}\equiv s\bar{s}$.
Different possible orbital-flavor-spin-color configurations of the four-quark
subsystems in the five-quark system are numbered by $i$; $n_{r}$ and $l$ denote the inner radial and
orbital quantum numbers, respectively. 
The coefficients $C_{in_{r}l}$ in Eq.~(\ref{wfn}) can be related to the 
coupling between the valence three-quark and the corresponding five-quark components.
\begin{equation}
 C_{in_{r}l}=\frac{\langle QQQ(Q\bar{Q}),i,n_{r},l|\hat{T}|QQQ\rangle}{M_{B}-E_{in_{r}l}}\,,
\end{equation}
where $M_{B}$ is the mass of baryon B, $E_{in_{r}l}$ the energy of the relevant five-quark component,
and $\hat{T}$ a model dependent coupling operator.

The matrix elements of $\hat{T}$ between the three- and five-quark configurations are derived
using a $^{3}P_{0}$ version for the transition coupling operator
\begin{equation}
 \hat{T}=-\gamma\sum_{j}\mathcal{F}_{j,5}^{00}\mathcal{C}_{j,5}^{00}C_{OFSC}
\left [\sum_{m}
\langle1,m;1,-m|00\rangle\chi^{1,m}_{j,5}
\mathcal{Y}^{1,-m}_{j,5}(\vec{p}_{j}-\vec{p}_{5})b^{\dag}(\vec{p}_{j})d^{\dag}(\vec{p}_{5}) \right ]\,,
\label{op}
\end{equation}
where $\gamma$ is a dimensionless constant of the model, 
$\mathcal{F}_{i,5}^{00}$ and $\mathcal{C}_{i,5}^{00}$ 
denote the flavor and color singlet of the quark-antiquark pair $Q_{i}\bar{Q}$ in the five-quark 
system, and $C_{OFSC}$ is an operator to calculate the orbital-flavor-spin-color overlap between 
the residual three-quark configuration in the five-quark system and the valence three quark system.

The derived matrix elements $T$ for the 34 five-quark configurations show
that only 17 configurations, corresponding to the orbital quantum number $l=1$ and radial quantum number 
$n_r=0$, survive and matrix elements $T$ for all other ones vanish.

The probability of the sea quark in each baryon $B$ reads
\begin{equation}
P_B^{Q\bar{Q}}= \frac{1}{\mathcal{N}}
\sum_{i=1}^{17}\Bigg[ \Big ( \frac{T_i^{Q\bar{Q}} }{M_B-E_i^{Q\bar{Q}}} \Big )^2 \Bigg ],
\label{prob}
\end{equation}
with the normalization factor 
\begin{equation}
\mathcal{N}  \equiv  1+ \sum_{i=1}^{17} \mathcal{N}_i  
= 1+\sum_{i=1}^{17} \Bigg[ \Big ( \frac{T_i^{u\bar{u}}}{M_B-E_i^{u\bar{u}}}\Big )^2 +
\Big (\frac{T_i^{d\bar{d}}}{M_B-E_i^{d\bar{d}}}\Big )^2 +
\Big ( \frac{T_i^{s\bar{s}}}{M_B-E_i^{s\bar{s}}} \Big )^2 \Bigg ].
\label{norm}
\end{equation}
Notice that in Eq. (\ref{norm}) the first term is due to the valence three-quark states, 
while the second term comes from the five-quark mixtures.

Sigma terms are in general expressed via the strangeness fraction $y_N$ and the non-singlet component
$\hat {\sigma}$ defined as
\begin{eqnarray}
y_N&=& \frac{2 \langle p | s\bar{s} |p \rangle}
{\langle p | u\bar{u} + d\bar{d} |p \rangle}=
 \frac{2P_{s\bar{s}}}
{3+2( P_{u\bar{u}} + P_{d\bar{d}} ) },\\
\label{yn}
\hat {\sigma}&=& \hat {m} \langle p | u\bar{u} + d\bar{d} - 2 s\bar{s} |p \rangle.
\label{shat}
\end{eqnarray}
Then, the $\sigma$-terms read as follows:
\begin{eqnarray}
\sigma_{\pi N}&=& \frac{\hat {\sigma}} {1-2y_N},\\
\label{pi}
%
\sigma_{K N}^{I=0}&=& \frac{\hat {m} + m_s} {4\hat {m}}(1+2y_N)\sigma_{\pi N}, \\
\label{K}
%
\sigma_{\eta N}&=& \frac{\hat {m} + 2 {y}_N m_s} {3\hat {m}} \sigma_{\pi N},
\label{eta}
\end{eqnarray}
where $\hat {m} =(m_u + m_d)/2$, with $m_u$, $m_d$ and $m_s$ current quark masses.
 
%
\section{Results and discussion}
\label{Res}
%
In Table~(1) our results for probabilities (Eq. \ref{prob}) of $q \bar{q}$ ($q \equiv u, ~d$) and 
$s \bar{s}$ are given for each of the 17, non-vanishing, five-quark configurations.
Then, predictions of our model for the sea content of the octet baryons, in particle basis, are reported 
in Table~(2).
Notice small discrepancies between the values in the above Tables and those in Tables (VI) and (VII) in our
recent paper~\cite{An:2012kj}, due to a misprint with respect to the i=12 configuration numbers.
The only significant change concrns the $d \bar d$ component in $\Xi ^\circ$, the probability of which increases from 12.1\% to 13.8\%.

Finally, using Eqs.~(2.6)-(2.8), we extract values for the meson-nucleon $\sigma$-terms; 
namely, $\sigma_{\pi N}, \sigma_{K N}$ and $\sigma_{\eta N}$ (Table~(3)).
%
%
\begin{table}[t!]
\begin{center}
\caption{\footnotesize Predictions for probabilities of different five-quark components in the nucleon, 
$\Lambda$, $\Sigma$ and $\Xi$.
Upper and lower panels are for the configurations with $L_{4q}=1$ and $L_{4q}=0$, respectively. 
}
\begin{tabular}{rlccccc}
\\[-10PT]
\hline
\hline   \\ [-14PT]
N$^\circ$ &  Configuration  & Sea flavor &   N    &  $\Lambda $ & $\Sigma$  & $\Xi$ \\
\hline  \\ [-14PT]
1 & $[4]_{FS}[22]_{F}[22]_{S}$ & $q\bar{q}$ &  0.146  & 0.115  & 0.066  & 0.081  \\
                           
& & $s\bar{s}$ &  0.010 & 0 & 0.020 & 0     \\
                         
2 & $[4]_{FS}[31]^{1}_{F}[31]_{S}$ & $q\bar{q}$ &    0.073  & 0   & 0.054 & 0.028  \\

& & $s\bar{s}$ &  0  & 0        & 0.009  & 0.016 \\

3 & $[4]_{FS}[31]^{2}_{F}[31]_{S}$ & $q\bar{q}$ &    0 & 0.052 & 0.003 &  0.006 \\

& & $s\bar{s}$ &   0.006 & 0.013 & 0  & 0   \\

4 & $[31]_{FS}[211]_{F}[22]_{S}$ & $q\bar{q}$ & 0 & 0.003& 0.011 & 0.010  \\                           

& & $s\bar{s}$ & 0.004 & 0.003 & 0   & 0   \\

5 & $[31]_{FS}[211]_{F}[31]_{S}$ &  $q\bar{q}$ &  0 & 0.002  & 0.009 &  0.008 \\

& & $s\bar{s}$ &   0.003 & 0.002 & 0  & 0   \\

6 & $[31]_{FS}[22]_{F}[31]_{S}$ & $q\bar{q}$ & 0.007 & 0.011 & 0.004 & 0.012 \\

& & $s\bar{s}$ &   0.002 & 0   & 0.005 & 0   \\

7 & $[31]_{FS}[31]_{F}^{1}[22]_{S}$ & $q\bar{q}$ & 0.018 & 0 & 0.018 &  0.011  \\

& & $s\bar{s}$ &  0  & 0   & 0.004 & 0.009 \\
                          
8 & $[31]_{FS}[31]_{F}^{2}[22]_{S}$ 
& $q\bar{q}$ &  0  & 0.016 & 0.001  &  0.002  \\

& & $s\bar{s}$ & 0.003 & 0.006 & 0   & 0  \\
                          
9 & $[31]_{FS}[31]_{F}^{1}[31]_{S}$ & $q\bar{q}$ &  0.005 & 0 &  0.006 &  0.003 \\

& & $s\bar{s}$ &   0  & 0   & 0.001   & 0.003 \\
                          
10 & $[31]_{FS}[31]_{F}^{2}[31]_{S}$ 
& $q\bar{q}$ &  0 &  0.004 & 0  & 0.001 \\
                          
& & $s\bar{s}$ & 0.001  & 0.002  & 0  & 0 \\
                          
\hline


11 & $[31]_{FS}[211]_{F}[22]_{S}$ & $q\bar{q}$ & 0   &   0.006 &   0.022 & 0.021 \\

& & $s\bar{s}$ & 0.009 & 0.006 & 0   & 0 \\

12 & $[31]_{FS}[211]_{F}[31]_{S}$& $q\bar{q}$ & 0  &  0.005 &  0.019 & 0.018 \\

& & $s\bar{s}$ & 0.008 & 0.006 & 0  & 0  \\

13 & $[31]_{FS}[22]_{F}[31]_{S}$ & $q\bar{q}$ & 0.015 &  0.025 &  0.010 & 0.028 \\

& & $s\bar{s}$ & 0.004 & 0  & 0.011 & 0     \\

14 & $[31]_{FS}[31]_{F}^{1}[22]_{S}$ & $q\bar{q}$ &  0.042 & 0  & 0.043 &  0.026 \\
                          
& & $s\bar{s}$ &  0  & 0  & 0.010 & 0.022 \\

15 & $[31]_{FS}[31]_{F}^{2}[22]_{S}$ 
& $q\bar{q}$ &  0  &  0.037 &  0.002 & 0.005 \\                                
                          
& & $s\bar{s}$ &  0.007 & 0.015 & 0  & 0 \\

16 & $[31]_{FS}[31]_{F}^{1}[31]_{S}$ & $q\bar{q}$ &  0.012 & 0  & 0.013 &  0.008 \\                                
                          
& & $s\bar{s}$ &  0   & 0  & 0.003 & 0.007 \\

17 & $[31]_{FS}[31]_{F}^{2}[31]_{S}$ & $q\bar{q}$ & 0  &  0.010 &  0  &  0.001 \\                                
                          
& & $s\bar{s}$ &  0.002   & 0.004 &  0   & 0   \\

\hline
\hline
                              
\end{tabular}
\end{center}
\label{nump}
\end{table}

The pion-nucleon $\sigma$-term is the most studied one and the often quoted value was 
reported in Ref.~\cite{Gasser:1990ce}: $\sigma_{\pi N}$=45$\pm$8 MeV, which was obtained from
$\pi N$ data analysis with taking into account the current algebra result generated by the 
quark masses. As it can be inferred from Table~(3), various results reported in literature
agree with the canonical value within 2$\sigma$, and our result falls at the lower band 
within 1$\sigma$.

In Table~(3) results from other authors are also shown, coming from Chiral Constituent Quark
Models ($\chi CQM$)~\cite{Dahiya:2011uh}, Perturbative Chiral Constituent Quark 
(P $\chi CQ$)~\cite{Inoue:2003bk}, 
Lattice QCD ($LQCD$)~\cite{Durr:2011mp,Bali:2011ks,Shanahan:2012wh}, and Chiral Perturbation Theory 
($\chi PT$)~\cite{Alarcon:2011zs}.

%
\begin{table}[t!]
\begin{center}
\caption{\footnotesize Predictions for the sea content of the octet baryons.  
} 
\begin{tabular}{lcccc}
\\[-10PT]
\hline
\hline  \\ [-10PT]
  Baryon   &   $ \bar{u}$      & $\bar{d}$   & $\bar{s}$ & $\bar{u}+\bar{d}+\bar{s}$ \\
\hline  \\ [-10PT]
p            & 0.100 & 0.218 & 0.058 & 0.376  \\
$\Lambda$    & 0.143 & 0.143 & 0.058 & 0.344  \\
$\Sigma^{+}$ & 0.101 & 0.179 & 0.063 & 0.343 \\
$\Sigma^{0}$ & 0.140 & 0.140 & 0.063 & 0.343  \\       
$\Xi^{0}$    & 0.131 & 0.138 & 0.057 & 0.326  \\
\hline
\hline
\end{tabular}
\end{center}
\label{sea}
\end{table}
\begin{table}[ht]
\begin{center}
\caption{\footnotesize Predictions for the meson-nucleon $\sigma$-terms (in MeV), with  
$\hat {\sigma}$=35 MeV (Ref.~\cite{Gasser:1990ce}) and $m_s / \hat {m}$=25 (Ref.~\cite{Leutwyler:1996qg}).}

\begin{tabular}{llccc}
\\[-10PT]
\hline  
\hline  \\ [-10PT]
  Reference   &   Approach & $\sigma_{\pi N}$ & $\sigma_{K N}$ & $\sigma_{\eta N}$\\
\hline  \\ [-10PT]
Present work & $E\chi CQM$ & 37 & 256 &  32 \\
Dahiya {\it et al.}~\cite{Dahiya:2011uh} & $\chi CQM$ & 31 & 196 &  31 \\
Inoue {\it et al.}~\cite{Inoue:2003bk} & $P \chi CQ$ & 55 & 33 &  96 \\
Durr {\it et al.}~\cite{Durr:2011mp} & $LQCD$ & 39$\pm$4 &  &   \\
Bali {\it et al.}~\cite{Bali:2011ks} & $LQCD$ & 38$\pm$12 &  &   \\
Shanahan {\it et al.}~\cite{Shanahan:2012wh} & $LQCD$ & 45$\pm$6 & 300$\pm$40 & \\
%
Alarcon {\it et al.}~\cite{Alarcon:2011zs} & $\chi PT$ & 43$\pm$5 &  &   \\ 
Alarcon {\it et al.}~\cite{Alarcon:2011zs} & $\chi PT$ & 59$\pm$7 &  &   \\
\hline
\hline
\end{tabular}
\end{center}
\label{sigma}
\end{table}

Our result for $\sigma_{\pi N}$ is close enough to all other predictions, except in the case
of a recent $\chi PT$ approach~\cite{Alarcon:2011zs}, where two different partial wave analysis 
(PWA) lead to: 43$\pm$5 MeV and 59$\pm$7 MeV; only the former being compatible with ours. 
It is worth noting that the smaller value is obtained using the PWA of the Karlsruhe-Helsinki
group~\cite{Koch:1982pu} and the larger one via a more recent PWA provided by 
SAID~\cite{Arndt:2006bf}. 

In Ref.~\cite{Dahiya:2011uh} it is reported that non-relativistic quark model underestimates all
three meson-nucleon $\sigma$-terms, and $\chi CQM$ under exact $SU(3)$ symmetry overestimates them.
Actually, taking into account the $SU(3)$ symmetry breaking, results come out very close to those found
within the present work. That is also the case for the extracted $\sigma_{K N}$ and
$\sigma_{\eta N}$.
  
Results from a $P \chi CQ$~\cite{Inoue:2003bk} lead to $\sigma$-terms, all three in disagreement
with our findings. 

For $\sigma_{K N}$, our result is compatible with those reported in 
Refs.~\cite{Shanahan:2012wh,Dahiya:2011uh}.

Finally, lattice QCD calculations~\cite{Durr:2011mp,Bali:2011ks,Shanahan:2012wh} are producing 
results mainly for $\sigma_{\pi N}$, with including strangeness content of the baryons and find 
strange $\sigma$-term to be in the range 20 - 50 MeV, though with rather large 
uncertainties~\cite{Shanahan:2012wh}. For $\sigma_{\pi N}$, $LQCD$ approaches favor values
significantly smaller than those obtained via dispersion relations, using the
SAID $\pi N$ phase-shift analysis~\cite{Arndt:2003if}, and giving~\cite{Hite:2005tg} $\sigma_{\pi N}$=81$\pm$6.
%
\section{Summary and conclusions}
\label{Sum}
An extended chiral constituent quark model, embodying genuine five-quark mixture in the ground state baryon octet
wave functions, was briefly presented, underlining the sea quark content.
We put forward predictions of our complete model for the percentage, per flavor, of the sea quark content for $N$, 
$\Lambda$, $\Sigma$ and $\Xi$.

Those predictions were then used to calculate the meson-nucleon $\sigma$-terms and led to: 
$\sigma_{\pi N}$=37 MeV, $\sigma_{K N}$=256 MeV and $\sigma_{\eta N}$=32 MeV.
Our findings compare satisfactorily with results from recent studies based on chiral
constituent quark model~\cite{Dahiya:2011uh} and lattice QCD~\cite{Durr:2011mp,Bali:2011ks,Shanahan:2012wh}, 
while showing significant discrepancies with values determined via perturbative chiral quark 
approach~\cite{Inoue:2003bk} and dispersion relations~\cite{Hite:2005tg}.
 
Extension of the present work to predict $\sigma$-terms for meson-$\Lambda$, meson-$\Sigma$ and 
meson-$\Xi$ is in progress. 

\end{document}